\def\simlt{\lower.5ex\hbox{$\; \buildrel < \over \sim \;$}}
\def\simgt{\lower.5ex\hbox{$\; \buildrel > \over \sim \;$}}

\RequirePackage{lineno} 
\documentclass[preprint2]{emulateapj}
\usepackage{amsmath}

\slugcomment{submitted to the ApJL}
\shorttitle{Atmospheric Circulation of Brown Dwarfs}
\shortauthors{Zhang and Showman}
\begin{document}
\nocite{*}

\title{Atmospheric Circulation of Brown Dwarfs: Jets, Vortices, and Time Variability}
\author{Xi Zhang and Adam P. Showman}
\affil{Department of Planetary Sciences and Lunar and Planetary Laboratory, University of Arizona, AZ 85721}
\email{Contact email: xiz@lpl.arizona.edu}

\begin{abstract}

A variety of observational evidence demonstrates that brown dwarfs
exhibit active atmospheric circulations. In this study we use a
shallow-water model to investigate the global atmospheric dynamics in
the stratified layer overlying the convective zone on these rapidly
rotating objects.  We show that the existence and properties of the
atmospheric circulation crucially depend on key parameters including
the energy injection rate and radiative timescale. Under conditions of
strong internal heat flux and weak radiative dissipation, a banded
flow pattern comprising east-west jet streams spontaneously emerges
from the interaction of atmospheric turbulence with the planetary
rotation.  In contrast, when the internal heat flux is weak and/or
radiative dissipation is strong, turbulence injected into the
atmosphere damps before it can self-organize into jets, leading to a
flow dominated by transient eddies and isotropic turbulence instead.  
The simulation results are not very sensitive to the form of the forcing.
Based on the location of the transition between jet-dominated 
and eddy-dominated regimes, we suggest that many brown dwarfs may
exhibit atmospheric circulations dominated by eddies and turbulence
(rather than jets) due to the strong radiative damping on these
worlds, but a jet structure is also possible under some realistic conditions.
Our simulated light curves capture important features from observed
infrared lightcurves of brown dwarfs, including amplitude variations
of a few percent and shapes that fluctuate between single-peak and
multi-peak structures.  More broadly, our work shows that the
shallow-water system provides a useful tool to illuminate fundamental
aspects of the dynamics on these worlds.

\end{abstract}


\section{Introduction}

Brown dwarfs are characterized by a
vigorously convective interior overlain by a stratified weather layer
(Burrows et al. 2006). Increasing observational evidence, including
chemical disequilibrium (e.g., Saumon et al. 2006; Leggett et
al. 2007), condensates and clouds (e.g., Tsuji 2002; Knapp et
al. 2004), infrared flux variability (e.g., Artigau et al. 2009; 
Radigan et al. 2012; Apai
et al. 2013), and surface patchiness (Crossfield et al.~2014) imply
the existence of strong vertical mixing and large-scale atmospheric
motion on brown dwarfs. To understand the atmospheric convection and
circulation, several models have been put forward. A local-box
simulation by Freytag et al. (2010) showed that convection could
trigger gravity waves in the stratified layer, causing vertical mixing
and helping maintain thin dust clouds. Showman \& Kaspi (2013)
presented the first three-dimensional (3D) general circulation model
of brown dwarfs and demonstrated that the atmosphere and interior
dynamics are rotationally dominated.  However, many aspects of
the expected atmospheric dynamics remain uncertain. Especially, 
there is a lack of global simulations investigating the stratified layer on brown dwarfs.

The surface patchiness suggested in the observations raises major questions
about the nature of the atmospheric circulation on brown dwarfs.
In particular, are the flows on brown dwarfs zonally banded, consisting
of numerous east-west jet streams like those on Jupiter and Saturn?  Or
do they consist primarily of turbulence and eddies with no
preferred directionality? The issue has major implications
for interpretations of variability.

To answer these questions, we introduce an idealized global
shallow-water model to study the atmospheric circulation on brown
dwarfs.  The shallow water model has been extensively used in
the atmospheric studies on Earth (e.g., Sadourny
1974), giant planets (e.g., Cho and Polvani 1996;
Showman 2007) and exoplanets (Showman \& Polvani 2011). Ours
are the first global numerical simulations of the dynamics in 
the stratified atmospheres of brown dwarfs.

\section{Shallow Water Model and the Brown Dwarf Regime}

We adopt a two-layer model, with constant densities in each layer.
The upper layer, of lower density, represents the flow in the
stratified atmosphere (pressure less than several bars), and the lower
layer, of greater density, represents the convective interior.  The
interface between the layers represents an isentrope near the
radiative-convective boundary (RCB).  In the limit where the lower
layer is quiescent and infinitely deep, this system reduces to the
one-layer shallow-water equations for the flow in the upper layer
(e.g., Vallis 2006)
\begin{subequations}
\begin{align}
\frac{D\bold{u}}{Dt} + \nabla(gh) + f \bold{k} \times \bold{u} = \bold{R}\\
\frac{D(gh)}{Dt}+gh \nabla \cdot \bold{u} = g \frac{h_{\rm eq}-h}{\tau_{rad}} +S
\end{align}
\end{subequations}
where $\bold{u}$ is the horizontal velocity vector,
$f=2\Omega \sin\phi$ is Coriolis parameter, $\Omega$ is the
rotation rate, $\phi$ is latitude, $\bold{k}$ is the upward unit
vector, $gh$ is the geopotential, where $g$ is gravity and $h$ is the
layer thickness, and $\bold{R}$ is an advection term required to account
for momentum transport between the layers (Showman \& Polvani 2011).

The forcing term $S$ represents the exchanges of mass, such
as convective overshoot, across the RCB 
(see Section 2.2 below). Radiative heating and cooling are
represented by a Newtonian cooling scheme, which relaxes the layer
thickness to a specified, constant radiative-equilibrium thickness
$h_{\rm eq}$ on a timescale $\tau_{\rm rad}$.

Our brown dwarf shallow-water model is modified from the NCAR Spectral
Transform Shallow Water Model (STSWM, Hack \& Jakob 1992), which
solves the shallow-water equations using the spectral transform method
in spherical geometry on the entire globe. The spatial resolution is
~$0.7^\circ$ in longitude and latitude. A $\bigtriangledown^6$
hyperviscosity is used to maintain numerical stability. More details
are available in Hack \& Jakob (1992). The initial condition is a flat
layer of thickness $h_{\rm eq}$ with no winds. Next, we will
introduce several important parameters identifying the brown dwarf
regime.

\subsection{Deformation Radius and Layer Thickness}
The Rossby deformation radius ($L_R$) measures the horizontal length
scale within which the atmospheric flow is more significantly
influenced by buoyancy rather than rotation. In a three-dimensional
stratified atmosphere, $L_R=NH_p/f$, where $H_p=R_g T/\mu g$ is the
pressure scale height, $R_g$ is the gas constant, $T$ is temperature,
and $\mu$ is the mean molecular mass. $N=[g/T (g/c_p
+dT/dz)]^{1/2}=g[\gamma /(c_pT)]^{1/2}$ is the
Brunt-V$\ddot{\mathrm{a}}$is$\ddot{\mathrm{a}}$l$\ddot{\mathrm{a}}$
frequency of the atmosphere, where $\gamma=[1+(c_p/g) dT/dz]$ is a
measure of the subadiabaticity of the stratified atmosphere;
$\gamma=1$ for an isothermal atmosphere and $\gamma=0$ for a fully
adiabatic profile. $c_p$ is the specific heat of the atmosphere, and
$z$ is altitude. The observed rotation rates, surface gravities and
temperatures for L and T dwarfs range from $10^{-4}$ to $10^{-3}$
$\mathrm{s^{-1}}$ (Reiners \& Basri 2008), 500-3000 m
$\mathrm{s^{-2}}$ and 750-2200 K (Burrows et al. 2006), respectively.
Given the temperature profile of a typical brown dwarf with
temperature of 1000 K and surface gravity of 500 m $\mathrm{s^{-2}}$,
we estimate $\gamma\sim0.1$ above the RCB.  Provided the
rotation rate $\Omega \sim10^{-3}$ $\mathrm{s^{-1}}$, the deformation
radius is $L_R\sim$ 1000 km, similar to the Jupiter value (Vasavada \&
Showman 2005).

We specify the reference geopotential, $gh_{\rm eq}$, so
that the model's deformation radius matches the expected value.
In the shallow-water
system, $L_R = \sqrt{gh_{\rm eq}}/f$, implying that:
\begin{equation}
gh_{eq}=(f L_R)^2=\gamma \kappa^2 c_p T   
\end{equation}                  
where $\kappa=R/\mu c_p$ is $\sim$2/7 for a hydrogen-dominated
atmosphere. Inserting brown-dwarf parameters implies that
the appropriate value of $gh_{\rm eq}$ is
$10^5$--$10^6$ $\mathrm{m^2}$ $\mathrm{s^{-2}}$.

\subsection{Forcing Parameters}
\label{forcing}

The mass source term $S$ represents all possible mechanisms by which
convection perturbs the stratified layer, including perturbations of
the RCB itself by the convection, convective overshoot and mixing
across the RCB, and latent heat release due to condensation of
silicates and iron. Since there is no single estimate of $S$, we specify its functional form 
parametrically in space and time. The mean value of the spatial pattern is zero 
and the standard deviation is $s_m$. Analogy with the isentropic-coordinate
primitive equations (Vallis, 2006) implies that forcing amplitude
$s_m$ relates to a three-dimensional atmosphere via
\begin{equation}
s_m \approx gh {q\over \Delta\theta}
\end{equation}
where $\Delta\theta$ is the difference in potential temperature across
the weather layer ($\sim$$10^2\,$K for a layer one scale height thick
given the expected stratifications) and $q$ is the net heating rate
at the RCB.  Given the
velocities expected by mixing length theory 
(Showman \& Kaspi 2013), convective overshoot
should produce deviations from the radiative-equilibrium temperature
just above the RCB of $\sim$1--$10\rm\,K$, which for radiative time
constants of $10^5\rm\,s$ implies local, spatially varying heating
rates induced by convection at the RCB of
$10^{-5}$--$10^{-4}\rm\,K\,s^{-1}$.  Given $gh\sim
10^6\rm\,m^2\,s^{-2}$, we expect $s_m \sim
0.1$--$1\rm\,m^2\,s^{-3}$.

We explore two schemes for the
convective forcing.  In the first, we assume that convection perturbs
the stratified layer everywhere simultaneously and continuously in
time with a small characteristic horizontal wavelength. The forcing pattern 
is specified by a superposition of spherical harmonics within the range 
of total spatial wavenumbers $n = n_f - \Delta n$ to $n_f+\Delta n$ and 
zonal wavenumbers between $-n$ and $n$, where $n_f$ 
is the central wavenumber of the forcing.
In this study, we adopt $n_f=40$ and $\Delta n=2$, based on the spatial 
pattern of temperature maps from 3D simulations 
(Showman \& Kaspi 2013). At each time step, the amplitude of each 
spherical harmonic is randomly 
 varied between 0 and $s_m$  and the orientation of each spherical 
 harmonic is randomly varied (with a pole position of the coordinate 
 system in which the spherical harmonic is evaluated that is random 
 in longitude and latitude).  These are summed into a forcing 
function $F(\lambda,\phi)$ at longitude($\lambda$) and latitude($\phi$). 
The standard deviation of $F$ is then normalized to $s_m$ at each time step.
The forcing pattern, $S(\lambda, \phi, t)$, is marched forward in
time as a Markov process: 
\begin{equation}
S(\lambda,\phi,t+\delta t) = (1-\alpha^2)^{1/2}S(\lambda,\phi,t) + \alpha F(\lambda,\phi)
\end{equation}
with a de-correlation factor $\alpha=\delta t/\tau_s$, where $\delta
t$ is the timestep, $\tau_s$ is the forcing timescale. 

In the second scheme, we postulate that the convection occurs
episodically over a small fraction of the horizontal area, as is
typical of moist convection on Earth and Jupiter.  In this case,
we randomly introduce isolated mass pulses that are circular Gaussians 
on the sphere with a radius $r_{s}$ and forcing timescale $\tau_{s}$. 
The forcing pattern at time $t$ associated with a given storm, $s_{\rm storm}(\mathbf{r},t)$, is:
\begin{equation}
s_{\rm storm}(\mathbf{r}, t) = s_m\mathrm{exp}\left[-\frac{|\mathbf{r}-\mathbf{r_0}|^2}{r_{s}^2}-\frac{(t-t_0)^2}{\tau_{s}^2}\right]
\end{equation}
where $s_m$ is the maximum forcing amplitude occurring 
at time $t_0$, $|\mathbf{r}-\mathbf{r_0}|$ is the distance from the convection center $\mathbf{r_0}$ to any given 
point $\mathbf{r}$ on the sphere.   The contributions of different
storms, each occurring at different locations $\mathbf{r_0}$ and times
$t_0$, are summed to obtain the total forcing, $S(\mathbf{r,t})=
\sum s_{\rm storm}(\mathbf{r},t)$.  The mass 
pulses are introduced randomly over the globe and in time (see Showman 2007 for details).

The forcing timescale is uncertain. Under typical brown dwarf
conditions, convective transport times across a scale height are
$\sim$$10^3\rm\,s$, but the convection (and the mixing and wave
breaking resulting from it above the RCB) may organize on longer
timescales, much as cumulus convection in Earth's tropics organizes
across a wide range of spatial and temporal scales (e.g., Mapes et al. 2006; Kiladis et al. 2009).
We consider $\tau_s \sim 10^3-10^5$ s.

Energetically, the forcing drives the circulation by generating
available potential energy\footnote{APE is the difference in potential
energy between the stratified atmosphere and a reference state
(Vallis, 2006).} (APE), which is converted to kinetic energy
associated with large-scale winds. The production rate of
APE due to a single gaussian mass pulse averaged over space and cell
lifetime is $\sim \pi^2 \tau_s r_s^2 s_m^2/4$ (Showman 2007). For a
globally convective system like a brown dwarf atmosphere, we estimate
$\sim A\tau_s s_m^2$, where $A$ is the global surface
area. We define the energy injection rate ($\epsilon$) as the 
ratio of the APE production rate to the total potential energy 
of the equilibrium state, $\sim A(gh_{eq})^2/2$. 
With our estimated forcing parameters, the energy injection 
rate on a typical Jupiter-size brown
dwarf is $\sim 10^{-8}-10^{-5}$ $\mathrm{s^{-1}}$. By comparison, 
$\epsilon$ of Jovian system is $\sim 10^{-11}$ $\mathrm{s^{-1}}$.

\subsection{Radiative Relaxation}
 
The radiative relaxation timescale ($\tau_{rad}$) is estimated as:
\begin{equation}
\tau_{rad} \sim \frac{p}{g} \frac{c_p}{4 \sigma T^3}  
\end{equation}
where $p$ is the pressure of the typical IR photosphere level (optical depth
$\sim$1). For a typical brown dwarf of $\sim$1000 K, the radiative
timescale is $\sim10^5$ s. Since
the effective temperature of brown dwarfs ranges from several hundred
(e.g., Y type) to several thousand kelvins(e.g., L0 type), the
radiative timescale can vary by several orders of magnitude
($\sim10^4-10^7$ s). By comparison, the radiative relaxation timescale
for Jupiter is $\sim 10^9$ s. 
Radiative cooling might be also important for the cloud formation on 
brown dwarfs (Helling et al. 2001).

\section{Jets versus Isotropic Turbulence on Brown Dwarfs}

\begin{figure}[b]
 \centering \includegraphics[width=0.45\textwidth]{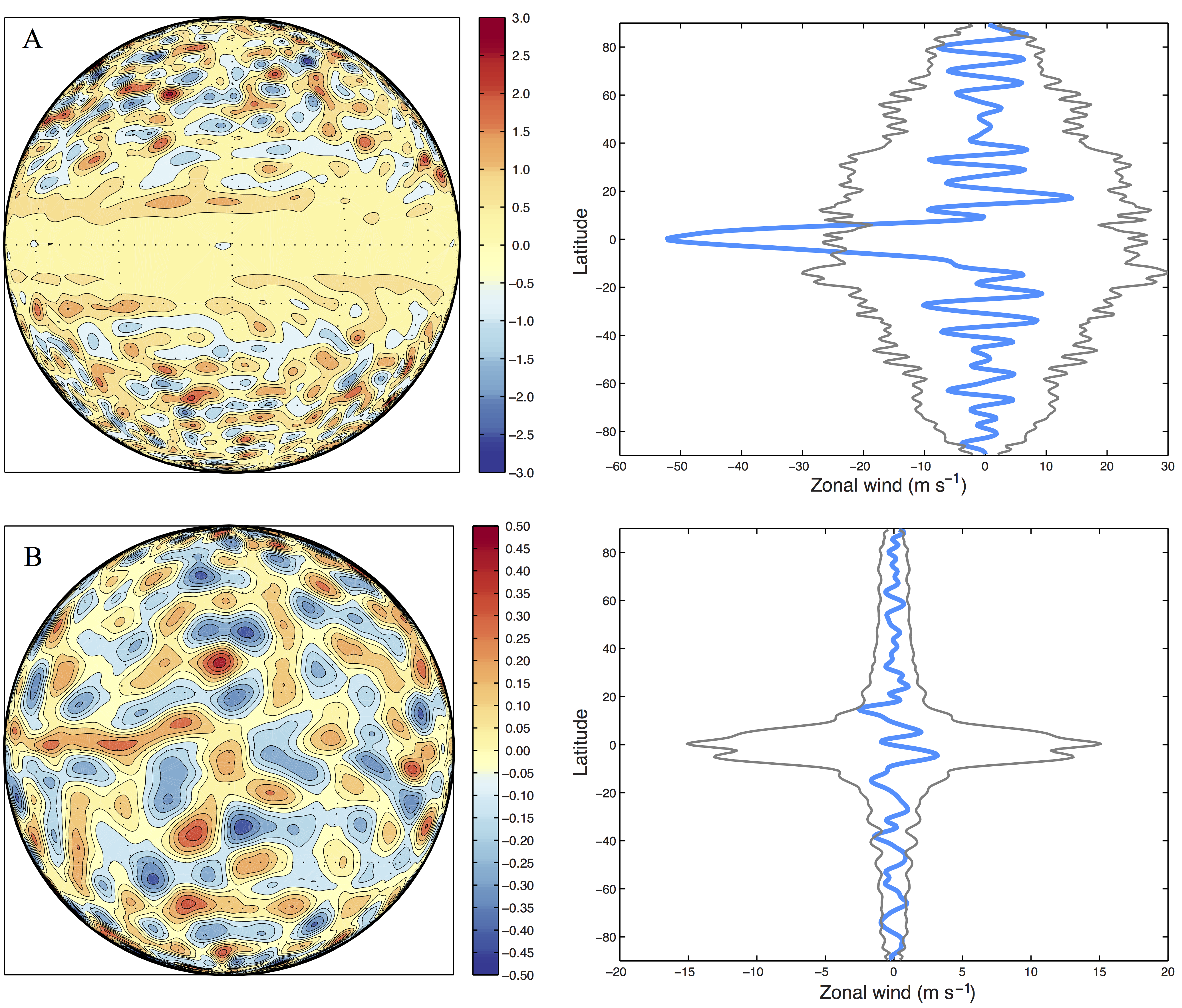} \caption{Two
    atmospheric regimes of brown dwarfs ($\Omega \sim10^{-3}\mathrm{s^{-1}}$) under
   forcing wavenumber 40. Upper: jet case (case A); lower: isotropic turbulence 
   case (case B). Left: geopotential
    anomaly map (in units of $10^5$ $\mathrm{m^2}$ $\mathrm{s^{-2}}$);
    right: zonal-mean zonal wind (blue) and standard deviation in
    longitude of the zonal wind at each latitude (gray). Both models
    have a small deformation radius of $\sim$$10^3\rm\,$km.}
\end{figure}

We find two major regimes of behavior depending on the strengths of
the forcing and dissipation. For strong energy injection and weak
dissipation, large-scale east-west (zonal) jet streams\footnote{
i.e., zonal mean zonal wind that exceeds the standard deviation of
zonal wind in longitude} emerge, particularly at the equator, where
they dominate the flow.  In contrast, for weak energy injection and
strong dissipation, jets are absent, and eddies and turbulence
dominate the dynamics.  Here we show two idealized simulations with
$n_f=40$, with a rotation period of 1.6 hours and an equilibrium geopotential of $5\times 10^5$
$\mathrm{m^2}$ $\mathrm{s^{-2}}$ (Fig. 1). Case A, corresponding to a
cold brown dwarf atmosphere ($\tau_{\rm rad}\sim 10^7$ s) with 
$\tau_s=10^5$ s and $s_m=0.1 \rm\,m^2\,s^{-3}$, exhibits a strong equatorial westward jet
with peak wind velocity of 50 m $\mathrm{s^{-1}}$ and long-lived
vortices in the middle and high latitudes. Case B corresponds to a
hotter brown dwarf atmosphere ($\tau_{\rm rad}\sim 10^5$ s) with
$\tau_s=10^3$ s and $s_m=0.1 \rm\,m^2\,s^{-3}$, implying that
the energy injection rate is 100 times weaker and the dissipation is
100 times stronger than in case A.  As a result, transient eddies and
isotropic turbulence, rather than jets, dominate the entire
globe. The geopotential variation in case A is almost a factor of 10
larger than that in case B. Over the ranges of rotation rate
($10^{-3}-10^{-4}$ $\mathrm{s^{-1}}$) and spectral wavenumber (20--70)
we explored, the system shows similar behavior.

{\begin{figure}[t]
  \centering \includegraphics[width=0.45\textwidth]{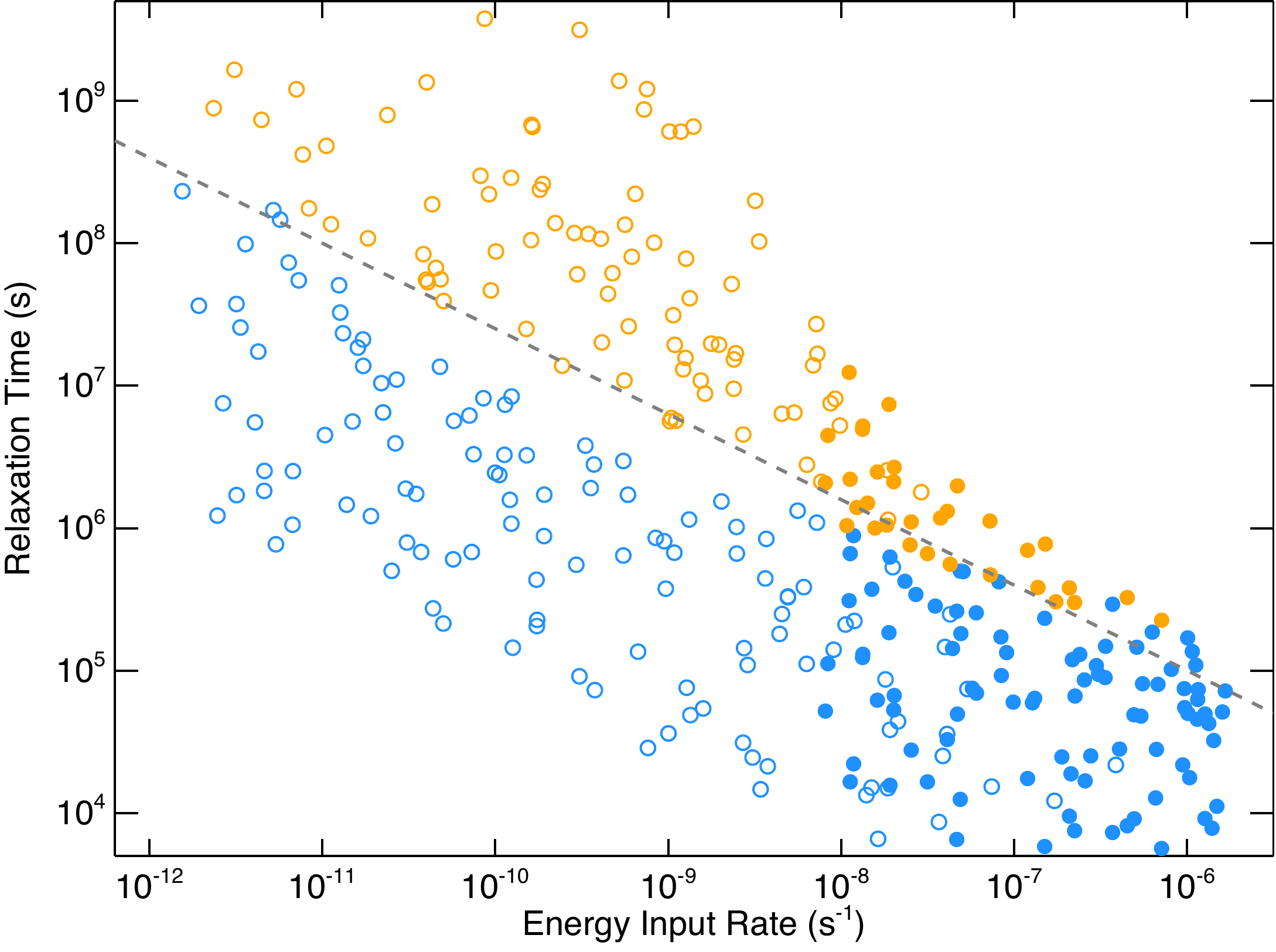} \caption{Regime
    diagram showing models exhibiting zonal jets
    (orange) and those without jets (blue) as function
    of energy input rate and relaxation timescale. Open circles: cases
    with local pulse forcing; filled circles: cases with global
    spectral forcing.}
\end{figure}

To systematically investigate the dichotomy of jet- versus turbulence-dominated, 
we performed several hundred simulations over a wide range of forcing and
dissipation parameters. Atmospheres with an equilibrium geopotential
of $10^5$ $\mathrm{m^2}$ $\mathrm{s^{-2}}$ were forced by
either local mass pulses with randomly-chosen size, strength, and lifetime,
or global spectral forcing with spherical harmonics ($n_f=40$) and random
forcing amplitude.  In the $\epsilon-\tau$
coordinate plane (Fig.~2), all cases fall into two distinct regimes:
with or without jets. The two regimes are separated by a boundary
line with slope of $\sim$0.6. The same trend emerges in
both the local pulse cases and global spectral cases. Therefore we
conclude that the emergence of jet structure in a shallow water system
is largely controlled by the energy injection rate and dissipation
timescale. Given the values in Sections 2, 
colder giant planets like Jupiter might be located in the upper-left corner 
of Fig. 2 and brown dwarfs are probably in the lower-right corner.

Why jets emerge in turbulent, rotating stratified atmospheres remains
incompletely understood.  In an intrinsically 2D system approximating 
the stratified layer, when forcing occurs at small scales, it is generally 
expected that energy transfers to large scales via an ``inverse energy 
cascade'' (Vallis 2006).  In the presence of damping,
large coherent structures such as long-lived vortices and strong jets
can develop only if energy can be transferred to large scales before it
is dissipated.  Two associated scales might be important. The first
one is the ``halting scale'', associated with the peak wavenumber
($k_h$) of the equilibrated kinetic energy (or available potential
energy) spectrum. The second one is the ``Rhines scale'',
$k_\beta \sim(\beta/U_{\rm rms})^{1/2}$ (Rhines 1975), a
characteristic scale for significant east-west banding to occur, where
$\beta$ is the latitudinal gradient of the Coriolis parameter and
$U_{\rm rms}$ is the root mean square of the zonal wind
velocity. Previous studies (e.g., Marcus et al. 2000; Danilov \&
Gurarie 2001; Smith et al. 2002), most of which are non-divergent and
hence significantly more idealized than the shallow-water system, have
generally found that the necessary condition for zonal jets to emerge
is $k_h<k_\beta$.  Qualitatively, a similar criterion seems likely in
the shallow-water system.

Interestingly, even the cases with equatorial jets are largely dominated 
by long-lived vortices at mid-to-high latitudes (Fig.~1).  This likely results from
the small deformation radius in these simulations, which is known to
suppress the Rhines effect and jet formation under appropriate
conditions (Okuno and Masuda 2003, Showman 2007).  To test this
hypothesis, we performed several simulation with thick layers and
deformation radii up to half a planetary radius, which, when damping
is sufficiently weak, indeed exhibit the formation of multiple jets
both at the equator and at high latitude (Fig.~3). As 3D
atmospheres exhibit multiple modes, including a baroclinic mode of
small deformation radius and a barotropic mode of large deformation
radius, it is possible that the atmospheric circulation on brown
dwarfs with sufficiently long radiative time constants will
qualitatively resemble a superposition of Fig.~3 and the top
panel of Fig.~1.

\begin{figure}[t]
  \centering \includegraphics[width=0.44\textwidth]{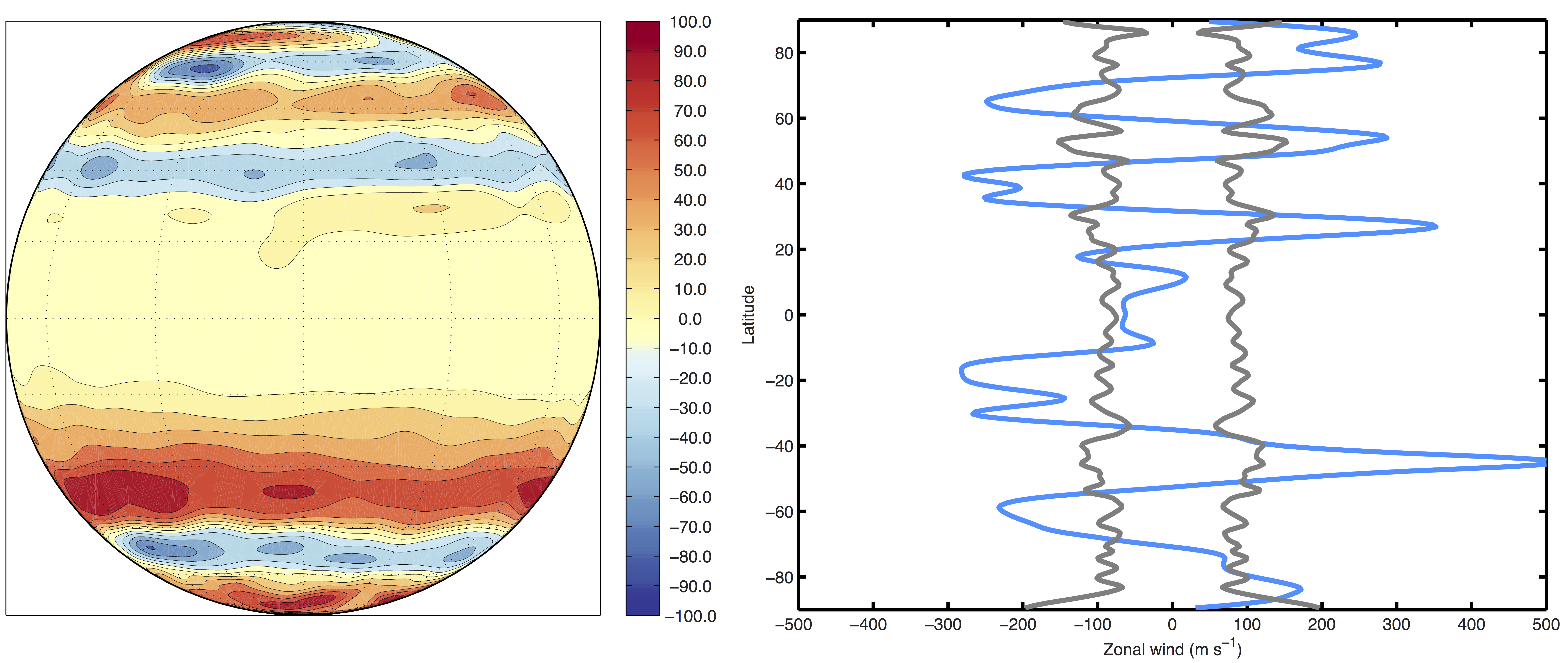} \caption{A
    brown dwarf ($\Omega \sim10^{-3}$ s) simulation with $n_f=40$, $gh_{eq}=5\times 10^8$ $\mathrm{m^2}$
    $\mathrm{s^{-2}}$, $\tau_{\rm rad}\sim10^7$ s, $\tau_s \sim10^5$ s, 
    and $s_m=10^2 \rm\,m^2\,s^{-3}$. Left:
    geopotential anomaly map (in units of $10^5$ $\mathrm{m^2}$
   $\mathrm{s^{-2}}$); right: zonal-mean zonal wind (blue) and
    standard deviation in longitude of the zonal wind at each latitude
    (gray). This model has a large deformation radius of 0.3 planetary radii.}
\end{figure}

\begin{figure*}
     \includegraphics[width=1.0\textwidth]{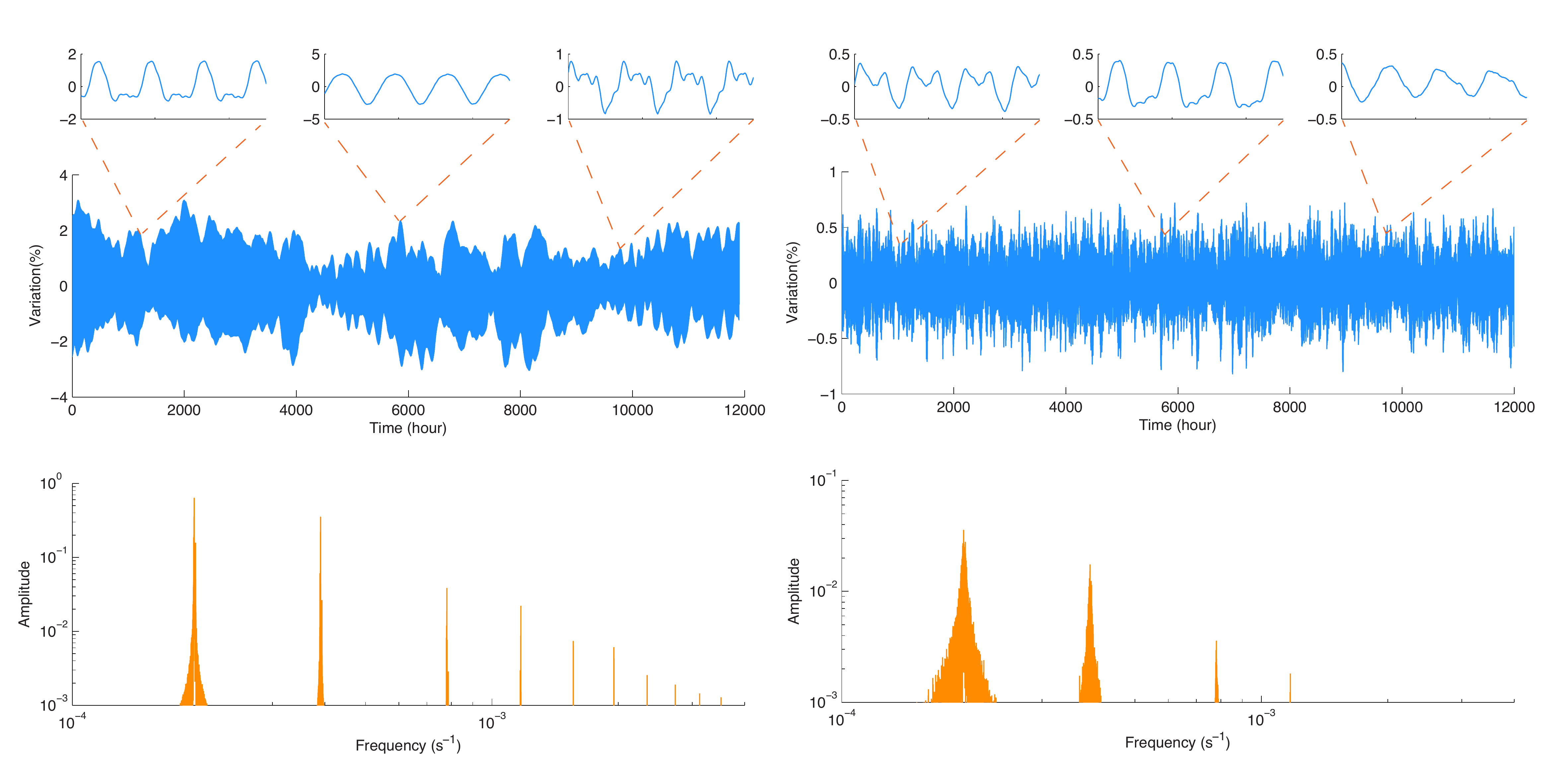}   
     \caption{Light curve analysis for case A (left) and case B (right). Upper: time variability of face-on geopotential anomaly, with three zoom-in charts; Lower: Single-sided amplitude spectra via FFT. Only noise exists at frequencies shorter than $10^{-4} \rm\,s^{-1}$.}
\end{figure*}

\section{Implications for Light Curves}

We use the global shallow-water model to diagnose the long-term system
behavior of brown dwarf atmospheres. Both the short-term and long-term
behavior of our simulations, from hours to years, can help illuminate
current and upcoming infrared light curve observations over various
timescales (e.g., Artigau et al. 2009; Apai et al. 2013).  
As a proxy of the emergent flux, the height field $\bar{h}(t)$ averaged over the
face-on hemisphere as a function of time $t$ can be obtained via:
\begin{equation}
\bar{h}(t)=\int^{\pi/2+\Omega t}_{-\pi/2+\Omega t} \int^{\pi/2}_{-\pi/2}h(\phi,\lambda)R^2\cos^2 \phi \cos\lambda d\phi d\lambda 
\end{equation}
In Fig.~4 we illustrate the time variations of geopotential anomalies of 
cases A and B (Fig. 1) after the systems reach statistically steady states. The
geopotential anomalies are normalized to the equilibrium geopotential.

On multi-hour timescales, the simulated light curves are dominated by
the rotation period, as atmospheric features rotate into or out of the
Earth-facing hemisphere. The amplitude and shape of light curves
remain similar over several rotation periods, but change
dramatically over longer timescales (tens to hundreds of rotation
periods).  Frequently, the light curves exhibit double-peak or
triple-peak features, qualitatively explaining the current observed
light curve shapes (e.g., Artigau et al. 2009). 
The amplitude of the case without jet (case B) is less than the jet
case (case A). The latter reaches up to several percent, similar to
the observed flux variations (e.g., Artigau et al. 2009). 
The peak-to-peak amplitude of the fluctuations evolves on 
timescales of tens (case B) to hundreds (case A) of hours, shown by the blue
envelope.  
In Fig.~1, the fractional variations in local height are order-unity,
which is similar to the fractional variations in effective temperature
between cloud and less-cloudy regions inferred from observations
(Artigau et al. 2009, Radigan et al. 2012).  Thus, if the patchy
cloud structure on a brown dwarf had the same geometry as in 
our simulations, it should also produce lightcurve variations
of $\sim$1\% in spectral windows such as $J$.

A spectral analysis for both cases via Fast Fourier Transform (FFT)
technique showed that the dominant mode is the rotational mode with
a period of $\sim$1.6 hours (Fig. ~4, bottom row), and the other
modes are the (even) multiples of the rotational mode as a fundamental
frequency.   This analysis suggests that, although the dynamics
itself exhibits a range of inherent frequencies, the rotational modulation
dominates the lightcurves.

\section{Conclusion} 

We demonstrated using a global shallow-water model that zonal
(east-west) jet streams and local vortices can form under conditions
appropriate to brown dwarfs. The existence and properties of the
jets crucially depend on the radiative damping timescale and the
rate at which convection injects energy into the atmosphere. Strong
internal heat flux (i.e., strong convective forcing) and weak
radiative dissipation favors the formation of large-scale jets in the
atmosphere, as shown in our simulations, whereas a weaker forcing and
stronger dissipation will halt the formation of zonal jets and favor
the generation of (horizontally) isotropic turbulence.

The radiative timescale for most brown dwarfs above 1000 K is
$\sim10^5$ s or less, and longer with colder atmospheres. The rate 
of energy injection at the RCB is less certain, which depends on the forcing
mechanisms, either from the dry and moist convective processes, or
from the upward propagating atmospheric waves. A crude estimate of a
typical brown dwarf implies that its energy injection rate is several
orders of magnitude larger than that on Jupiter. 

Our simulations suggest that many hot brown dwarfs may be
eddy/turbulence dominated rather than jet dominated at typical IR
photosphere levels. If a brown dwarf is cold enough, such as Y dwarfs,
or the damping timescale is long, or the forcing is strong enough, 
zonal jets should form, as seen on Jupiter.  Nevertheless, it is difficult to
precisely relate our forcing and damping parameters to the brown dwarf
spectral sequence, as $\epsilon$ and $\tau_{\rm rad}$ will depend on
the detailed nature of the convection, the thickness of the layer
communicating with the photosphere (which may include the top portion
of the convection zone), and other factors.  While our finding of a
regime transition is robust and should carry over to the full 3D system, these uncertainties imply that more
realistic three-dimensional models will be necessary to pin down
precisely where the transition falls in the brown dwarf spectral
sequence.  Despite these uncertainties, it is clear that Jupiter is in the
jet-dominated regime, providing some confirmation of our findings.

The long-time integration of the shallow water system provides a new
tool to understand the observed light curve variations. We found that
our simulated brown dwarf atmospheres are dominated by the rotational modulation
in short-term light curves, with lightcurve shapes that vary from
single to multi-peaked periodic structures and amplitudes of a few
percent, qualitatively consistent with recent observed infrared flux
variations.

\section{Acknowledgements}		
This research was supported by Bisgrove Scholar Program to X. Z. and NSF grant AST1313444 to A.P.S. The authors wish to thank the International Summer Institute for Modeling in Astrophysics (ISIMA) where part of the model was developed.

\end{document}